\documentclass[twocolumn, manuscript]{revtex4}
\usepackage{graphicx}
\begin{document}

\title{Energy diffusion in strongly driven quantum chaotic
systems: Role of correlations of the matrix elements}
\author{P.V. Elyutin and A.N. Rubtsov}
\email{pve@shg.phys.msu.su} \affiliation {Department of Physics,
Moscow State University, Moscow 119992, Russia}
\date{\today}

\begin{abstract}
The energy evolution of a quantum chaotic system under the
perturbation that harmonically depends on time is studied for the
case of large perturbation, in which the rate of transition
calculated from the Fermi golden rule (FGR) is about or exceeds
the frequency of perturbation.  For this case the models of
Hamiltonian with random non-correlated matrix elements demonstrate
that the energy evolution retains its diffusive character, but the
rate of diffusion increases slower than the square of the
magnitude of perturbation, thus destroying the quantum-classical
correspondence for the energy diffusion and the energy absorption
in the classical limit $\hbar \to 0$. The numerical calculation
carried out for a model built from the first principles (the
quantum analog of the Pullen - Edmonds oscillator) demonstrates
that the evolving energy distribution, apart from the diffusive
component, contains a ballistic one with the energy dispersion
that is proportional to the square of time.  This component
originates from the chains of matrix elements with correlated
signs and vanishes if the signs of matrix elements are randomized.
The presence of the ballistic component formally extends the
applicability of the FGR to the non-perturbative domain and
restores the quantum-classical correspondence.

\vspace{10mm} PACS numbers {05.45.-a}
\end{abstract}
\maketitle

\section{Introduction}
The problem of susceptibility of chaotic  systems to perturbations
has attracted much attention in the last decade [1 - 9].  This
problem is fundamental, since it includes the determination of the
response of a material system to an imposed external
electromagnetic field, the setup that is typical for many
experiments.  Due to the sensitivity of classical phase
trajectories or quantum energy spectra and stationary
wavefunctions of chaotic systems to small changes of their
parameters, the problem is challengingly difficult.  A consistent
and noncontroversial picture covering (albeit qualitatively) all
the essential cases of the response hasn't been yet drawn at
present. From the point of view of general theory, the problem is
related to the applicability of the concept of quantum-classical
correspondence to chaotic systems, that is a long-standing
question in its own right \cite{E88,+E88}.

We shall study a one-particle system with the Hamiltonian of the
form $\hat H = \hat H_0  - F\hat x\cos \omega t$, where $\hat H_0
\left( {{\bf{\hat p}},{\bf{\hat r}}} \right)$ is the Hamiltonian
of the unperturbed system; $\bf{\hat p}$  and $\bf{\hat r}$ are
the operators of Cartesian components of the momentum and of the
position of the particle.  The classical system with the
Hamiltonian function $H_0 \left( {{\bf{p}},{\bf{r}}} \right)$ will
be assumed to be strongly chaotic, that is, nearly ergodic on the
energy surfaces in a wide range of the energy values, system with
$d \ge 2$ degrees of freedom.  In the perturbation operator $\hat
V\left( t \right) = - F\hat x\cos \omega t$ the active variable
$\hat x$ is one of the Cartesian coordinates of the particle,
coupled to the external homogeneous force field.  The amplitude
$F$ in the following will be referred to as the field.  In the
following we shall deal with the quasiclassical case, when the
Planck constant $\hbar$ is small in comparison of the action scale
of the system $H_0$.

Under the influence of the perturbation the energy value
$E(t)=H_0(t)$ varies in a quasirandom way.  These variations
 can be frequently described as a process of the energy diffusion
\cite{LG91,J93}, when for the ensemble with the microcanonical
initial energy distribution $H_0(0)=E$ the dispersion of the
energy increases linearly with time, $\left\langle {\Delta E^2
\left( t \right)} \right\rangle  = 2Dt$, where $D\left(
{E,F,\omega} \right)$ is the energy diffusion coefficient.

If the external field $F$ is sufficiently small in comparison with
the appropriately averaged values of the forces acting on a
particle in the unperturbed system, then in the classical model
the energy diffusion coefficient $D$ can be expressed through the
characteristics of the unperturbed chaotic motion of the active
coordinate, namely
\begin{equation}\label{1}
D = \frac{\pi }{2}\omega ^2 F^2 S_x \left( {E,\omega } \right),
\end{equation}
where $S_x(E,\omega )$ is the power spectrum of the active
coordinate (the Fourier transform of its autocorrelation function)
for the motion on the surface with the constant energy value $E$
[9].    The same expression (1) in the case of weak perturbation
can be obtained in the classical limit from the quantum model. The
evolution of the quantum system can be treated as a sequence of
one-photon transitions between stationary states of the
unperturbed system $\left| n \right\rangle  \to \left| k
\right\rangle $, accompanied with the absorption or emission of
the quanta $\hbar \omega $. For small $\hbar$ the energy spectrum
of $\hat H_0 $ is quasicontinuous, thence the rates of transition
are given by the Fermi golden rule (FGR)
\begin{equation}\label{2}
\dot W_F  = \frac{\pi }{{2\hbar }}F^2 \left| {x_{nk} } \right|^2
\rho \left( {E_k } \right),
\end{equation}
where $x_{nk}$ is the matrix element of the active coordinate, and
$\rho(E_k)$ is the density of states near the final state of the
transition.  Although the matrix elements $x_{nk}$  in quantum
chaotic systems fluctuate wildly with the variation of $k$ [10,
11], the averaged squared quantity $\overline {\left| {x_{nk} }
\right|^2 } $ in the limit $\hbar  \to 0$ is smooth; it is
proportional to the power spectrum $S_x \left( {E,\omega }
\right)$ of the coordinate \cite{FP86,W87},
\begin{equation}\label{3}
\overline {\left| {x_{nk} } \right|^2 }  \approx \frac{{S_x \left(
{E,\omega } \right)}}{{\hbar \rho \left( E \right)}}.
\end{equation}
From Eqs. (2) and (3) we have for the transition rate
\begin{equation}\label{4}
\dot W_F  = \frac{\pi }{{2\hbar ^2 }}F^2 S_x \left( {E,\omega }
\right).
\end{equation}
Then for the energy dispersion for small $t$ we have $\left\langle
{\Delta E^2 } \right\rangle  = 2 D_Ft= 2\left( {\hbar \omega }
\right)^2 \dot W_F t$, that returns us to the Eq. (1) for the
energy diffusion coefficient.  It can be shown that the same
expression for $D$ holds also for large $t$ \cite{+E04}. The
energy absorption in chaotic systems comes as an epiphenomenon of
the energy diffusion [4].  With the account of the dependence on
the energy of the power spectrum $S_x \left( {E,\omega } \right)$
and the density of states $\rho \left( E \right)$ the diffusion
becomes biased, and the energy absorption rate $Q$ is given by the
formula \cite{+ESh96, C99}
\begin{equation}\label{5}
Q = \frac{1}{\rho }\frac{d}{{dE}}\left( {\rho D} \right).
\end{equation}

Although for weak fields $D$ does not depend on the Planck
constant $\hbar$, the condition of the applicability of Eq. (2)
does.  The FGR is, after all, only a formula of the first order
perturbation theory.  It is based on the assumption that the
transition process has a resonant character - that the width
$\Delta$ of the energy distribution of states populated from the
original one, given by the Weisskopf - Wigner formula $\Delta  =
\hbar \dot W$ \cite{WW30}, is small in comparison with the quanta
energy $\hbar \omega $.  From Eq. (4) it is evident that in the
classical limit $\hbar  \to 0$ this condition will be violated. In
the following we shall use the border value of the field $F_b$,
defined by the condition $\dot W_F \left( {F_b } \right) = \omega
_0 $, and refer to the domain $F \ge F_b $ as the range of the
strong field.

By analogy with other models, for strong fields one can expect a
slow-down of the growth of the energy diffusion coefficient $D$
and of the energy absorption rate $Q$.  For example, for a
two-level system with relaxation the quadratic dependence the
absorption rate $Q \propto F^2 $ for small field turns into a
field-independent value $Q_0$ for strong one.  The border is
determined by the condition  ${{\Omega ^2 } \mathord{\left/
{\vphantom {{\Omega ^2 } {\Gamma _1 \Gamma _2 }}} \right.
\kern-\nulldelimiterspace} {\Gamma _1 \Gamma _2 }}\sim 1$, where
$\Omega$ is the Rabi frequency and ${\Gamma _1, \Gamma _2 }$ are
longitudinal and transversal relaxation rates correspondingly
\cite{AE75}.  The rate of transitions from discrete to continuous
energy spectrum (that are basically covariant with the energy
absorption rate $Q$), studied in the context of the theory of
photoionization, for sufficiently strong fields can even decrease
with the increase of $F$ - the effect that is known as atom
stabilization by the strong field \cite{DK95}.

The slow-down of the energy diffusion in strong harmonic fields
for the model of quantum chaotic systems with random uncorrelated
matrix elements has been first demonstrated by Cohen and Kottos
\cite{CK00}.  A different approach \cite{+E06} has lead to
qualitatively the same results.  This slow-down destroys the
quantum-classical correspondence.

It has been demonstrated by Kottos and Cohen \cite{KC01} that for
the first-principles model that is constructed by the quantization
of the Hamiltonian of a classically chaotic system, the response
to a sudden change of the (otherwise) stationary Hamiltonian
measured by the energy spreading restores its classical behaviour
for sufficiently small values of $\hbar$ in contrast with the
model with random independent matrix elements.

The purpose of the present article is to study this phenomenon for
the harmonically driven system.

\section{Numerical experiment}

The system chosen is the Pullen - Edmonds oscillator \cite{PE81},
that describes the two-dimensional motion of a particle in the
quartic potential. The Hamiltonian of this system is
\begin{equation}\label{6}
H_0={1 \over {2m}}\left( {p_x^2+p_y^2} \right)+{{m\omega _0^2}
\over 2}\left( {x^2+y^2+{{x^2y^2} \over {\lambda ^2}}} \right).
\end{equation}
In the following we use the particle mass $m$, the frequency of
small oscillations $\omega_0$, and the nonlinearity length
$\lambda$ as unit scales, and write all equations in dimensionless
form.

The properties of chaotic motion of the Pullen - Edmonds model are
thoroughly studied \cite{M86,VZ87,EK89}.  With the increase of
energy the system becomes more chaotic both in extensive (that is
characterized by the measure of the chaotic component $\mu_s(E)$
on the surface of Poincare section) and in intensive (that is
measured by the magnitude of the Lyapunov exponent $\sigma (E)$)
aspects. For values of energy $E>2.1$ the measure $\mu_s >0.5$,
and chaos dominates in the phase space; for $E>5$ the chaotic
motion of the system is approximately ergodic \cite{M86}.

The matrix of the quantum Hamiltonian operator of the model Eq.
(6) has been calculated in the basis of the unperturbed
two-dimensional isotropic harmonic oscillator for the value $\hbar
= 0.05$.  Due to the symmetry of the system the submatrices with
different parities of the quantum numbers $n_x$ and $n_y$ can be
diagonalized separately.

By expanding the wavefunction of the system $\Psi \left({\bf{r}},
t \right)$ in the basis of the eigenstates $\left\{ {\varphi _m }
\right\}$ of the system $\hat H_0$
\begin{equation}\label{7}
\Psi \left({\bf{r}}, t \right) = \sum\limits_m {a_m (t) \varphi _m
\left( {\bf{r}} \right)\,} e^{ - i\omega _m t} ,
\end{equation}
we obtain for the amplitudes $a_m$ the system of equations
\begin{equation}\label{8}
i\frac{{da_k }}{{dt}} = \sum\limits_k {\Omega _{km} \cos \omega
t\,e^{i\omega _{km} t} a_m } ,
\end{equation}
where the quantities $\Omega _{kn}  = \hbar ^{ - 1} Fx_{kn}$ are
the Rabi frequencies of transitions.  This system has been solved
numerically for the 2352 amplitudes of the eigenstates with
"even-even" and "odd-even" parities of $n_x$ and $n_y$ that
include all states of these classes with energies in the band $10
\le E \le 12$.

For the initial conditions in the runs with different values of
$F$ and $\omega$ we have used the same normalized narrow
wavepacket with randomly chosen real amplitudes $a_k(0)$ with the
complexity (inverse participation ratio) $C = \left(
{\sum\limits_k {a_k^4 \left( 0 \right)} } \right)^{ - 1} = 21$,
the mean initial energy $\langle E \rangle = E_0 = 11.0$ and the
initial energy dispersion $\Delta E^2 \left( 0 \right) = 2.5
\times 10^{ - 3} $.

To expose the role of correlations of the matrix elements $x_{kn}$
on the energy kinetics we compare the properties of the "natural"
system with its "randomized" analog with the matrix elements
$y_{mn} = x_{mn}A_{mn}$, where $A_{mn}$ are the elements of a
symmetric matrix that take values 1 or -1 at random with equal
probabilities. This randomization, that has been introduced in
\cite{KC01}, destroys the correlation of matrix elements.  The
tests has shown that the results do not depend on the specific
choice of the matrix $A_{mn}$ within the limits of an error of
about 1\%.

\begin{figure}
[!ht]
\includegraphics[width=0.9\columnwidth]{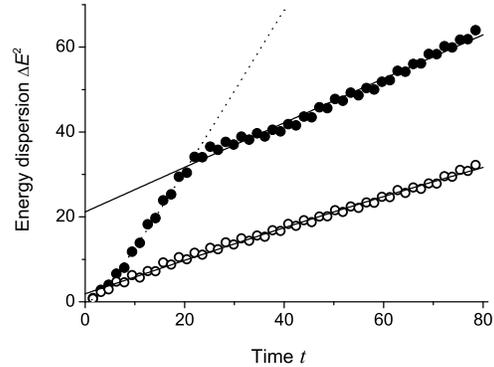}
\caption{\label{fig1} The dependence of the energy dispersion
$\Delta E^2$ on time $t$ for the natural (filled circles) and
randomized (open circles) models.  The energy dispersion is
measured in the units of $(\hbar \omega_0)^2$.  The field strength
corresponds to the Fermi point: $F=F_b=0.022$; the perturbation
frequency $\omega = 1.00.$}
\end{figure}

Figure 1 presents the typical dependence on time of the energy
dispersion for the natural (filled circles) and randomized (open
circles) models.  To extract the value of the energy diffusion
coefficient, the numerical dependence was fitted by the law of
evolution of the energy dispersion for the diffusion equation with
the constant $D$ on the interval $(-L, L)$ with impenetrable walls
on the borders and the initial condition in the form of the
$\delta (0)$ peak. This dependence can be described (with the
local accuracy better than 3\%) by the formula
\begin{eqnarray}\label{9}
 \Phi (a,b;t) = a\left( {1 - \exp \left( { - bt} \right)} \right) \times  \\
 \,\,\,\,\,\,\,\,\left( {1 + 0.633bt\exp \left( { - 1.161bt} \right)}
 \right)\nonumber
\end{eqnarray}
where $a = {{L^2 } \mathord{\left/ {\vphantom {{L^2 } 3}} \right.
\kern-\nulldelimiterspace} 3}$ and $b = {{6D} \mathord{\left/
 {\vphantom {{6D} {L^2 }}} \right. \kern-\nulldelimiterspace} {L^2
 }}$.  The best fits of the formula Eq. (9) with the numerical
 data are shown in the Fig. 1 by solid lines. The initial moment
 $t_0$ has been used as the third fitting parameter.

For once, it is clearly seen that the randomization suppresses the
process of the energy diffusion.  Secondly, for the natural model
one can see the presence of two different regimes - a fast initial
diffusion sharply slows down at a crossover time $t_c\simeq 20$.
We note that at this moment the energy dispersion $\Delta E^2
\left( {t_c } \right) = 32\left( {\hbar \omega _0 } \right)^2 $ is
much less than the saturated value $\Delta E_s^2 = a = 133\left(
{\hbar \omega _0 } \right)^2 $, that corresponds to the uniform
probability distribution throughout the band of the states taken
into account. To understand what happens at the crossover time we
have to study more closely the time development of the probability
distribution.

The overall form of the energy distribution for the natural model
is rather accurately approximated by the gaussian form that
follows from the model of the energy diffusion with the constant
$D$.  This agreement can be seen in Fig. 2, where the logarithm of
the energy density distribution is shown against the reduced
energy shift $\Delta \varepsilon = {{\left( {E - E_0 } \right)}
\mathord{\left/{\vphantom {{\left( {E - E_0 } \right)} {\hbar
\omega _0 }}} \right. \kern-\nulldelimiterspace} {\hbar \omega _0
}}$, where $E_0$ is the mean energy value of the initial
wavepacket.
\begin{figure}
[!ht]
\includegraphics[width=0.9\columnwidth]{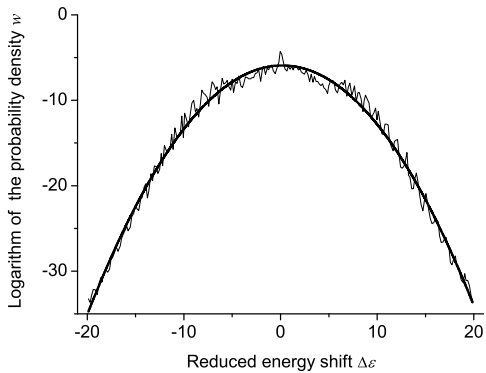}
\caption{\label{fig2} The logarithm of the energy density
distribution for the time $t=17.3$ for the parameters $F=0.606
F_b,\omega=1.00$. The grassy line is the numerical data, the solid
line presents the best fit of data with the parabola $p(\Delta
\varepsilon)=\alpha - \beta (\Delta \varepsilon)^2$.}
\end{figure}

Although the agreement seems to be very good, one must keep in
mind that the vertical scale of the graph is logarithmic.  By
subtraction of the parabolic fit from the numerical distribution
we come to the picture of deviations that is shown in the Fig. 3.

\begin{figure}
[!ht]
\includegraphics[width=0.9\columnwidth]{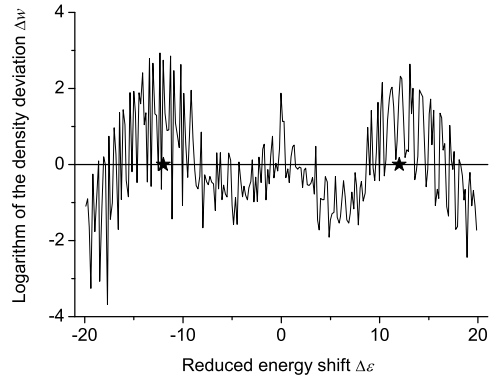}
\caption{\label{fig3} The deviation of the logarithm of the energy
density distribution from the best fitted parabola $\Delta w$  as
a function of the dimensionless energy deviation $\Delta
\varepsilon$ for the same parameters as in Fig.2. The black stars
note the maxima of the ballistic bumps (see the text)}.
\end{figure}

The peak at $\Delta \varepsilon = 0$ corresponds to the part of
the initial packet that is not depleted yet by the energy
spreading. Two bumps are clearly seen in the picture: their maxima
are located at $\Delta \varepsilon \simeq \pm 12$.  The
calculations show that these bumps propagate outwards with a
constant speed, that is ballistically.  The crossover time
corresponds to the moment when these bumps reach the borders of
the treated band of states.  Therefore, only the part of the graph
that precedes the $t_c$ corresponds to the properties of the
model; the second part is just an artifact of the calculation
scheme.  To estimate the diffusion coefficient we fitted the
dependence of the initial state by the two-parameter formula
\begin{equation}\label{10}
\phi \left( {D,\tau ;t} \right) = \frac{{2Dt^2 }}{{\tau  + t}} ,
\end{equation}
where the time shift $\tau$ accounts for the duration of the
initial stage, where the law of the dispersion growth is always
quadratic.  The best fitted function $\phi$ is plotted in  Fig. 1
by the dashed line.

The ballistic spreading of the energy distribution in the time
domain $t >  > {{2\pi } \mathord{\left/ {\vphantom {{2\pi } \omega
}} \right. \kern-\nulldelimiterspace} \omega }$ is well known for
the model of a one-dimensional resonantly excited harmonic
oscillator with the Hamiltonian $\hat H\left( t \right) = {{\left(
{\hat p^2  + \hat x^2 } \right)} \mathord{\left/ {\vphantom
{{\left( {\hat p^2  + \hat x^2 } \right)} 2}} \right.
\kern-\nulldelimiterspace} 2} - F\hat x\cos t$ . In the
quasiclassical domain (for large quantum numbers $n \gg 1$) the
matrix elements of the coordinate can be taken constant, $x_{nm} =
X\left( {\delta _{n,n - 1}  + \delta _{n,n + 1} } \right)$, where
$\delta_{ij}$ is the Kroneker delta - symbol.  With this
assumption in the rotating wave approximation for the initial
condition $a_k \left( 0 \right) = \delta _{kn} $ the probabilities
$w_k$ to find the system in the state $|k>$ are given by the
well-known formula
\begin{equation}\label{11}
w_k  =|a_k(t)|^2= J_{n - k}^2 \left( {\Omega t} \right),
\end{equation}
where $J_n(z)$ are the Bessel functions of the first kind and
$\Omega  = {{FX} \mathord{\left/ {\vphantom {{FX} {2\hbar }}}
\right. \kern-\nulldelimiterspace} {2\hbar }}$ is the value of the
Rabi frequency.  Equation (11) yields for the energy dispersion
\begin{equation}\label{12}
\Delta E^2 \left( t \right) = \frac{1}{2}\left( {\hbar \omega _0
\Omega t} \right)^2.
\end{equation}
However, the randomization of signs of matrix elements does not
influence the energy kinetics in this model.

The influence of randomization can be explained by the toy "double
ladder" model.  This system has the doubly degenerate equidistant
energy spectrum $E_n  = \hbar \omega _0 \left[ {{n \mathord{\left/
{\vphantom {n 2}} \right. \kern-\nulldelimiterspace} 2}} \right]$
where $[,]$ denotes the integer part of the number.  The matrix
elements of the coordinate connect each state $|n>$ to all four
states $|m>$ with the energy differences $E_n  - E_m  =  \pm \hbar
\omega _0 $:
\begin{equation}\label{13}
x_{nm}  = X\left( {\delta _{n,n - 2}  + \delta _{n,n - 1}  +
\delta _{n,n + 2}  + \delta _{n,n + 3} } \right)
\end{equation}
for even $n$ and
\begin{equation}\label{14}
x_{nm}  = X\left( {\delta _{n,n - 3}  + \delta _{n,n - 2}  +
\delta _{n,n + 1}  + \delta _{n,n + 2} } \right)
\end{equation}
for odd $n$.  In this model for the resonant perturbation $\hat
V\left( t \right) =-F\hat x\cos \omega _0 t$ the energy spreading
is ballistic, $\Delta E^2 \left( t \right) \propto t^2 $, whereas
the randomization of signs leads to the localization of the
quasienergy states, and the energy dispersion growth saturates.

One can suppose that the ballistic component in the quantum
chaotic model is carried through the subset of states that are
similar to the "double ladder" model.  The degree of correlation
of the matrix elements can be estimated from the construction
\begin{equation}\label{15}
T_n  = \sum\limits_{m,l,k} {x_{nm} x_{ml} x_{lk} x_{kn} },
\end{equation}
that describes the sum of contributions of all possible four
consequent transitions that start and end on the same state $|n>$.
The correlation index $\nu$ can be defined as the ratio of the
average value of $T$ for the randomized system to that of the
natural system.  For the "double ladder" model we have
\begin{equation}\label{16}
\nu  = \frac{{\left\langle {T_n^R } \right\rangle }}{{\left\langle
{T_n^N } \right\rangle }} = \frac{7}{{12}} = 0.583.
\end{equation}
The relatively large value of this number is explained by the
large contribution of symmetric contours with $m=k$, that are
invariant under the randomization.  For the Pullen - Edmonds model
the value of the correlation index $\nu = 0.72$ is rather close to
that of the "double ladder" model; that makes the analogy
plausible.

It must be stressed that the combined effect of the bulk diffusion
spreading and the overlaying ballistic bumps propagation produces
the \textit{linear} growth of the energy dispersion (see Fig.1)
that will be referred to as the effective diffusion.

We define the repression coefficient $R(F, \omega)$ as the ratio
of the effective energy diffusion coefficient to its value $D_F$
that follows from the FGR:

\begin{equation}\label{17}
R\left( {F,\omega } \right) = \frac{{D\left( {F,\omega }
\right)}}{{2\left( {\hbar \omega } \right)^2 \dot W_F }}.
\end{equation}
The dependence of the $R$ on the field strength is shown in Figs.
4 and 5 for two different values of the perturbation frequency.
\begin{figure}
[!ht]
\includegraphics[width=0.9\columnwidth]{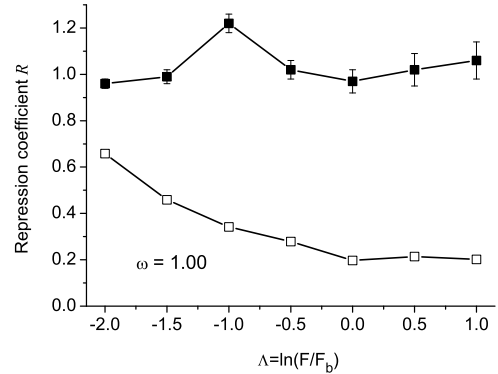}
\caption{\label{fig4} The dependence of the repression coefficient
$R$ on the logarithm of the field strength $\Lambda = \ln(F/F_b)$
for the frequency of perturbation $\omega = 1.00$.}
\end{figure}

\begin{figure}
[!ht]
\includegraphics[width=0.9\columnwidth]{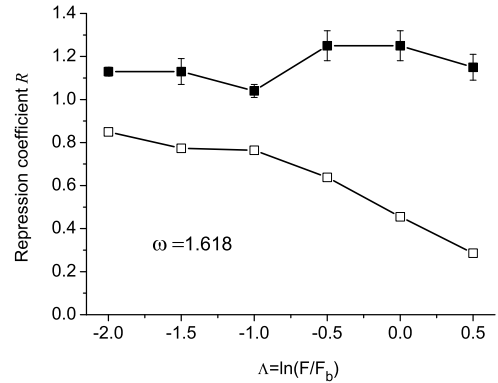}
\caption{\label{fig5} The dependence of the repression coefficient
$R$ on the logarithm of the field strength $\Lambda = \ln(F/F_b)$
for the frequency of perturbation $\omega = 1.62$.}
\end{figure}

For the natural model $R$ remains approximately constant with the
value close to unity, whereas for the randomized model the energy
diffusion slows down in qualitative agreement with the conclusions
of \cite {CK00, +E06}.

\section{Energy diffusion in the classical model}

The classical expression for the diffusion coefficient Eq. (1) is
derived in the limit of the infinitesimal perturbation, when one
can neglect the influence of the perturbation on the law of motion
of the active coordinate $x(t)$. Let's study the formation of this
coefficient.  We represent the external field in the form $F\sin
\omega t$ and denote by $t_n  = {{2\pi n} \mathord{\left/
{\vphantom {{2\pi n} \omega }} \right. \kern-\nulldelimiterspace}
\omega }$ the moments of time at which the external field take
zero values.  The variation of the energy for one field period
between these moments is exactly proportional to the field
strength,
\begin{equation}\label{18}
\Delta E_n  = F\int\limits_{t_n }^{t_{n + 1} } {\dot x\left( t
\right)} \sin \omega t\,dt = F\Delta _n .
\end{equation}
The quantities $\Delta_n$ we shall call the reduced variations of
the energy.  In the accepted approximation they do not depend on
the field strength.

The following calculations were carried out for the Pullen -
Edmonds oscillator on the energy surface $E=11.0$ and for the
perturbation frequency $\omega = 1.00$.

\begin{figure}
[!ht]
\includegraphics[width=0.9\columnwidth]{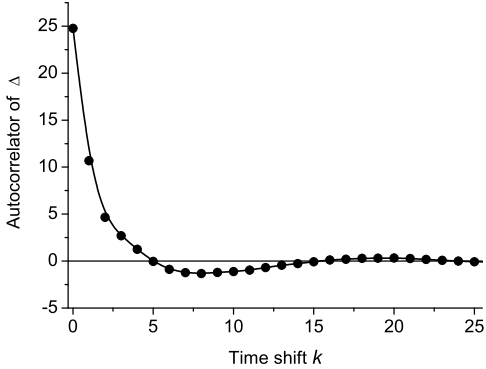}
\caption{\label{fig6} The dependence of the autocorrelator of the
reduced energy increments $B_\Delta  \left( k \right) =
\left\langle {\Delta _n \Delta _{n + k} } \right\rangle $ on the
time shift $k$ measured in periods of the perturbation on the
energy surface $E=11.0$ for the perturbation frequency $\omega
=1.00$.}
\end{figure}

The values of the reduced energy increments $\Delta_n$ on the
neighbouring time intervals are correlated.   Figure 6 presents
the form of the autocorrelation function of the reduced energy
increments. One can see that for the values of the time shift $k
\gtrsim 5$ the correlations become rather small.

The quantity
\begin{equation}\label{19}
d_K  = \frac{\omega }{{4\pi K}}\left( {\sum\limits_{i = 0}^{K - 1}
{\Delta _{n + i} } } \right)^2
\end{equation}
will be called the $K$-th approximant of the reduced diffusion
coefficient.  This is a proportionality coefficient between the
diffusion coefficient and the square  of the field amplitude,
calculated from the interval of time of $K$ consequent periods of
the field. The positive correlation of $\Delta_k$ for small k
produces the initial monotonous growth of the $d_k$ that rather
rapidly comes to a saturation.
\begin{figure}
[!ht]
\includegraphics[width=0.9\columnwidth]{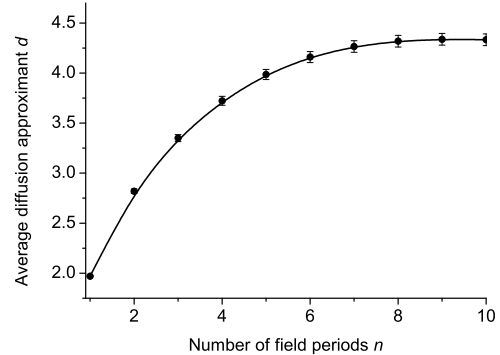}
\caption{\label{fig7} The dependence of the approximants of the
reduced diffusion coefficient $d_n$ on the duration of the time
interval measured in periods of the perturbation for the Pullen -
Edmonds model Eq. (6) on the energy surface $E=11.0$ for the
perturbation frequency $\omega =1.00$.}
\end{figure}

From the graph in  Fig. (7) it is  seen that already $d_8$ takes
the value that within the 1.5\% error margin is undistinguishable
from the asymptotic limit.  However, this average quantity is
formed by the contributions that differ by several orders of
magnitude. The graph in Fig. 8 shows the distribution of the
quantities $d_8$ in the log-log scale. The distribution is taken
from averaging over four ensembles of $10^5$ points each.

\begin{figure}
[!ht]
\includegraphics[width=0.9\columnwidth]{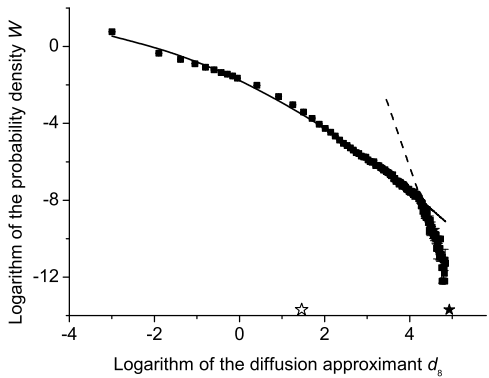}
\caption{\label{fig8} The distribution of the values of partial
contributions to the approximant $d_8$ from different points on
the trajectory of the Pullen - Edmonds model on the energy surface
$E=11.0$ for the perturbation frequency $\omega=1.00$. The thin
lines show the approximate forms of the distribution given by the
formulas in the text.  The black star marks the maximal partial
contribution $d_+$ that comes from the resonant trajectory, the
white star marks the position of the average value $<d_8>$. }
\end{figure}

The dominating part of this distribution is accurately fitted by
the dependence $\ln w = f(d) =  - 1.771 - 1.057\ln d - 0.095\left(
{\ln d} \right)^2 $ that is shown by the thin solid line.  For the
largest values of  $d_8$ another approximation is valid, $\ln w =
g(d) = 19.268 - 6.386\ln d$. This dependence is shown in Fig.8
with the thin dashed line.  In the domain of validity of the
approximation $f(d)$ the slope of the curve is less than unity:
the distribution is of the Zipf - Pareto type, in which the
dominating contribution to the average comes from the rare large
terms.  In our case the 20\% of the largest terms come with 82\%
contribution to the average. These large contributions come from
the bits of the trajectories in which the point oscillates almost
along the direction of the perturbing force nearly synchronously
with the perturbation. Theoretically the maximal value of $d_8$
originates from the motion with the law $x(t)=\sqrt{2 E}\sin (t)$
and equals to $d_+=4 \pi E/\omega=138.23$.

 In this resonant case the energy increment grows
linearly in time - that is, ballistically.

Thus we can indicate a classical counterpart to the quantum
dynamics of energy growth.  The quantum ballistic bumps are
analogous to the nearly resonant bits of the classical
trajectories with the quasiballistic energy increase.

\section{Conclusion}

By the numerical studies of the evolution of the energy
distribution in a harmonic external field in a system constructed
by the quantization of a classically chaotic Hamiltonian system,
thus retaining all correlations of the matrix elements, we have
found that the effective rate of the energy diffusion preserve its
quadratical dependence on the field strength on the domain of the
strong field, where the transition rate is comparable to the
perturbation frequency. In other words, the Fermi golden rule
appears to be valid far beyond the limits of the domain in which
its applicability can be justified.  This circumstance restores
the quantum - classical correspondence for the energy diffusion
and the energy absorption rate in the limit $\hbar \to 0$.

We have to admit that our studies are limited to a specific model,
studied for two values of the perturbation frequency.  However the
revealed mechanism of the ballistic component of the energy
distribution that propagates through the chains of matrix elements
with the correlated signs can be admitted as a universal,
especially with the account of the important contribution of the
quasiballistic parts of the trajectories to the energy diffusion
in the classical model. The more rigorous proof of the
universality demands further studies.

\section*{Acknowledgements}
The authors acknowledge the support by the "Russian Scientific
Schools" program (grant \# NSh - 4464.2006.2).

\end{document}